\documentclass[10pt,letterpaper,twocolumn]{article} 
\usepackage{ol2}
\usepackage[draft]{hyperref}
\usepackage{amsmath}

\begin{document}

\twocolumn[ 

\title{\textit{In situ} characterization of an optical cavity using atomic light shift}

\author{A. Bertoldi$^1$, S. Bernon$^1$, T. Vanderbruggen$^1$, A. Landragin$^2$, and P. Bouyer$^1$}
\address{$^1$Institut d'Optique, Univ. Paris Sud, CNRS, F--91127 Palaiseau, France}
\address{$^2$LNE-SYRTE, Observatoire de Paris, CNRS and UPMC, F--75014 Paris, France}

\begin{abstract}

We report the precise characterization of the optical potential obtained by injecting a
distributed--feedback erbium--doped fiber laser (DFB EDFL) at 1560 nm to the transversal modes of a folded
optical cavity.  The optical potential was mapped \textit{in situ} using cold rubidium atoms, whose
potential energy was spectrally resolved thanks to the strong differential light shift induced by the 1560
nm laser on the two levels of the probe transition. The optical potential obtained in the cavity is
suitable for trapping rubidium atoms, and eventually to achieve all--optical Bose--Einstein condensation
directly in the resonator.

\end{abstract}




]

\maketitle

Optical dipole traps \cite{grimm00} proved to be a reliable tool to manipulate ultracold neutral matter,
both for atoms and molecules. Once the detuning from the resonance is fixed, the depth of the optical
potential is directly proportional to the local intensity of the radiation. An optical cavity is a
straightforward way to increase the optical intensity, thanks to the long storage time of the photons in
the resonator. For this reason optical resonators are increasingly adopted in cold atom physics: after the
demonstration of cavity trapping of atoms \cite{mosk01,kruse03}, cavity cooling schemes have been proposed
\cite{vuletic00}, superradiance \cite{chan03} and collective atomic motion \cite{nagorny03} have been
observed in resonators. Recently, spin--squeezing of a cavity confined atomic sample was proved
\cite{schleier-smith09}, and Bose--Einstein condensates (BECs) have been coupled to the field of optical
resonators \cite{slama07,brennecke07}.

In this Letter, we report the characterization of the optical potential generated in a high finesse
optical cavity using cold $^{87}$Rb atoms as sensors of the local optical intensity. The scattering rate
on the D$_2$ line at 780 nm becomes strongly dependent on the local potential depth because of the
differential light shift of the two levels of the probe transition \cite{brantut08}, determined by the
1560 nm radiation pumping the cavity. The ring folded resonator was designed to have two of its beams
perpendicularly crossing at the center of the configuration. The optical potential obtained by locking the
DFB fiber laser to several transversal modes of the non-degenerate cavity was precisely characterized as
concerning shape and optical depth. The resulting configuration meets the requirements to optically trap
neutral atoms and could possibly bring to BEC directly in the optical resonator: the two arms consent a
large capture volume for atoms pre--cooled in a MOT, whereas the central crossing region provides a tight
confinement along all directions.

\begin{figure}[b]
\centerline{\includegraphics[width=6.0cm]{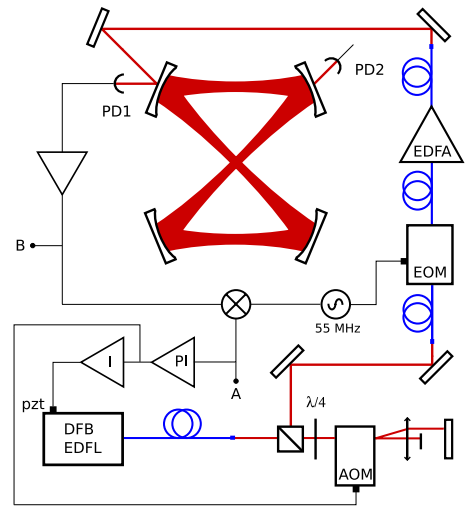}}
\caption{\label{fig:setup}(Color online) Frequency stabilization setup to lock the DFB EDFL to a
transversal mode of the folded optical cavity. In blue and red the fibered and the free space optical
path, in black the electronic connections. }
\end{figure}

An overview of the laser frequency stabilization system is presented in Fig. \ref{fig:setup}. The optical
cavity consists of four identical mirrors having a radius of curvature of 100 mm, and a dielectric coating
with reflectivity $R=0.99965$ at 1560 nm for \textit{p}--polarized light. The mirrors are mounted on a
rigid titanium platform and form a horizontal square with a diagonal of 90 mm. They are oriented to
produce a folded, ``8''--shaped cavity geometry. The resonator has a nearly concentric configuration, with
a free spectral range (FSR) of 976 MHz. Two mirrors are fixed, and one is mounted on a three axis
nanopositioning system providing more than one FSR of cavity tuning. The horizontal and vertical angles of
the fourth mirror are controlled by piezoelectric actuators, used to align the resonator. The cavity
assembly is tightly fixed on a CF 250 flange, which is mounted on the main ultra--high--vacuum chamber.

The optical resonator is pumped with the radiation produced by a single longitudinal mode DFB EDFL near
1560 nm (Koheras laser from NKT Photonics). The laser has a typical linewidth of a few kHz, an output power
of 100 mW, and a frequency noise spectral density dominated by $1/f$ components as reported in
\cite{kefelian09}. Before its injection in the resonator the laser radiation is amplified with a 5 W
erbium--doped fibered amplifier (EDFA with a gain of 37 dB, using a monomode polarization maintaining
fiber) to obtain an optical potential depth of the order of $k_B \times 1$ mK in each cavity arm for
rubidium atoms. The radiation is coupled to the resonator through a beam--expander and a tilted doublet to
optimize the mode--matching. The angle of the doublet allows to correct for the astigmatism of the cavity
modes, given by the off--axis incidence angle of 22.5$^{\circ}$ on the cavity mirrors. The degree of
astigmatism was calculated with ABCD matrix formalism for paraxial ray propagation \cite{kogelnik66}, and
experimentally confirmed by measuring the profile versus distance of the beam transmitted by one cavity
mirror.  A coupling efficiency of about 25\% was determined by measuring the reflected power on the input
mirror when scanning the cavity length across its TEM$_{00}$ resonance. The coupling efficiency of the
TEM$_{10}$ and TEM$_{20}$ modes was optimized by using phase masks, achieving 17\% and 12\%, respectively.
The scattering in the reverse mode at the cavity mirrors, determined by measuring the light transmitted in
the opposite direction, causes a 4\% lattice potential.

The laser is locked to a mode of the cavity using the Pound--Drever--Hall technique \cite{drever83}:
optical sidebands at about 55 MHz are generated using a fibered electro--optic modulator (EOM), and the
beatnote is detected in reflection with a InGaAs photodiode (PD1 in Fig. \ref{fig:setup}). The dispersive
signal obtained by demodulating the beatnote is used to lock the laser frequency to the cavity. A 250 kHz
bandwidth feedback is applied on an acousto--optic modulator (AOM) in double--pass through a proportional
and integral loop, whereas a 100 Hz feedback is applied to the piezo--electric element controlling the
laser cavity length. The main delay limiting the total correction bandwidth is the propagation time of the
RF signal in the AOM crystal from the electrode to the optical beam.

\begin{figure}[t]
\centerline{\includegraphics[width=8.5cm]{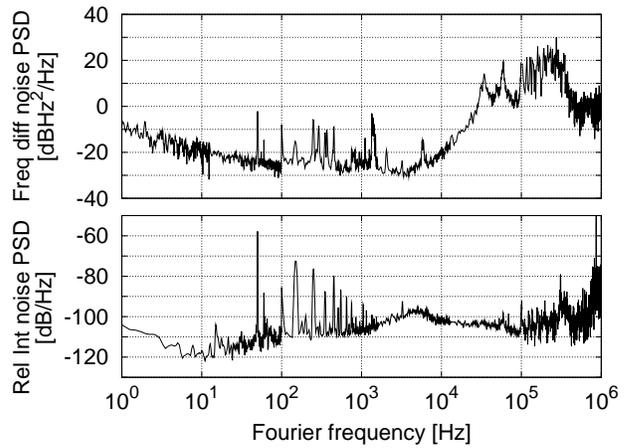}}
\caption{\label{fig:psd} Frequency difference noise (top) and relative intensity noise (bottom) versus
Fourier frequency of the fiber laser stabilized on the fundamental transversal mode of the optical
cavity.}
\end{figure}

The power spectral density (PSD) of the frequency difference noise between the laser and the cavity was
obtained by measuring the error signal for a closed servo loop (Fig. \ref{fig:psd}). For frequencies up to
100 kHz the signal was taken after the demodulation at point A in Fig. \ref{fig:setup} with a FFT spectrum
analyzer. For higher frequencies it was measured with a spectrum analyzer before the mixer at point B. The
noise spectral density shows the servo--bump of the fast AOM correction loop placed at about 250 kHz, and
stays below the 0.1 Hz/Hz$^{1/2}$ level between 10 Hz and 10 kHz. Below 100 Hz the frequency noise is
dominated by a $1/f$ component with 0.1 Hz/Hz$^{1/2}$ at 10 Hz. The intensity noise of the radiation field
in the cavity was measured evaluating the FFT of the transmission signal (PD2 in Fig. \ref{fig:setup}): as
shown on the spectrum of Fig. \ref{fig:psd}, the intensity noise PSD is always below 10$^{-9}$ Hz$^{-1}$
with respect to the DC level, except for narrow peaks in the acoustic band and close to 1 MHz. The
integral value of the noise PSD between 1 Hz and 10 kHz amounts to 2.8$\times$10$^{-6}$. 

In the crossing region of the two arms of the resonator was operated a $^{87}$Rb MOT. The absorption signal
of the laser cooled atoms was used to characterize the cavity. The radiation field at 1560 nm injected in
the cavity is close to the rubidium transitions 5P$_{3/2}$ -- 4D$_{3/2,5/2}$ at 1529 nm\footnote{The
two-photon transition on the D$_2$ line is inhibited by selection rules, and is further suppressed by
tuning the laser frequency so as to avoid the resonance condition.}: therefore it causes a widely stronger
red light shift of the upper level (5P$_{3/2}$) with respect to the ground one (5S$_{1/2}$) when the D$_2$
line is adopted to probe the atoms. More precisely, the shift ratio of the two levels at 1560 nm, given by
the scalar polarizability ratio, is 47.5. If the optical power in the cavity causes a light shift of the
5P$_{3/2}$ level much larger than the natural linewidth of the probe transition ($\Gamma$ = 2$\pi \times
$6.065(9) MHz), the potential energy of the atoms in the 1560 nm beam is spectrally resolved. Atoms at
different depths in the optical potential are thus imaged by changing the probe detuning with regard to the
atomic resonance \cite{brantut08}.

Using the absorption of a released MOT cloud, the optical resonator was aligned by tilting the two angles
of the movable cavity mirror. The two central arms of the cavity were vertically overlapped. A precise
orthogonality is important in relation with atom trapping, because it eliminates interferences between the
two crossing beams when the radiation polarization is set to be in the plane of the cavity.

\begin{figure}[t]
\centering
\includegraphics[width=2.78cm]{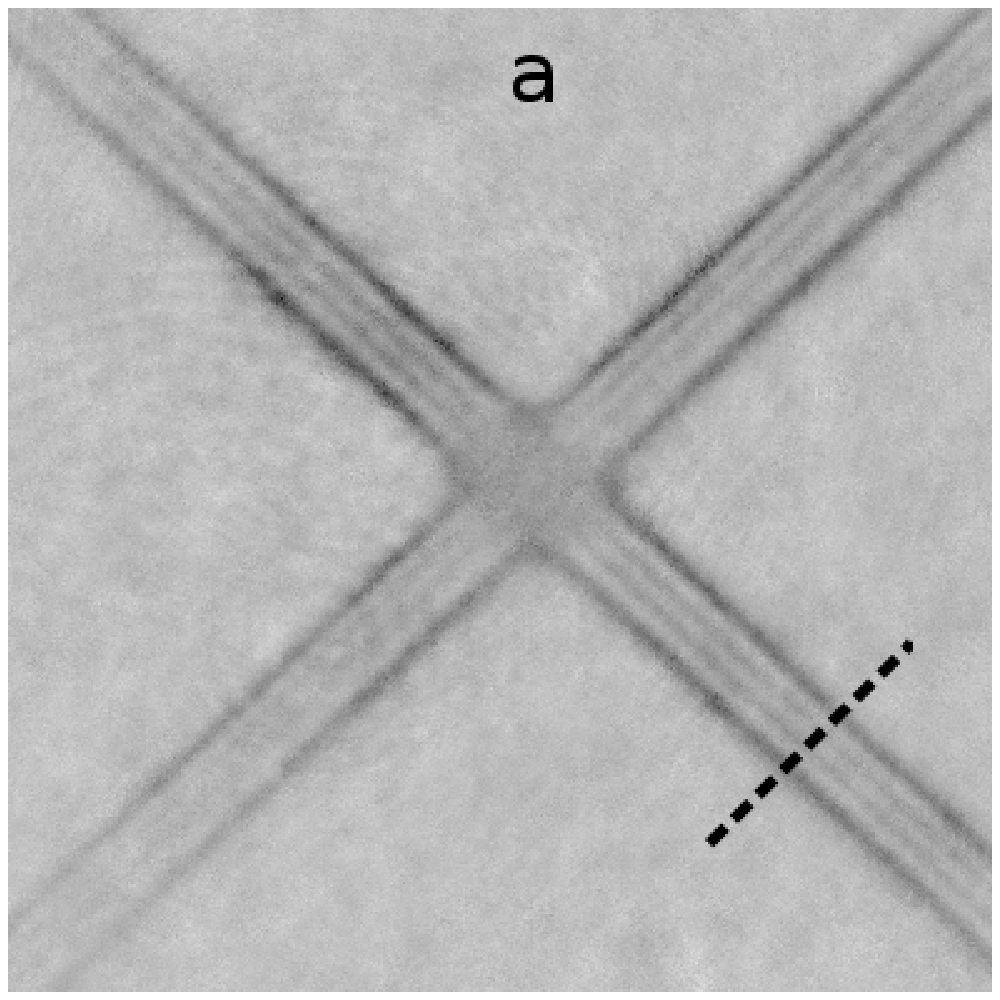}
\includegraphics[width=2.78cm]{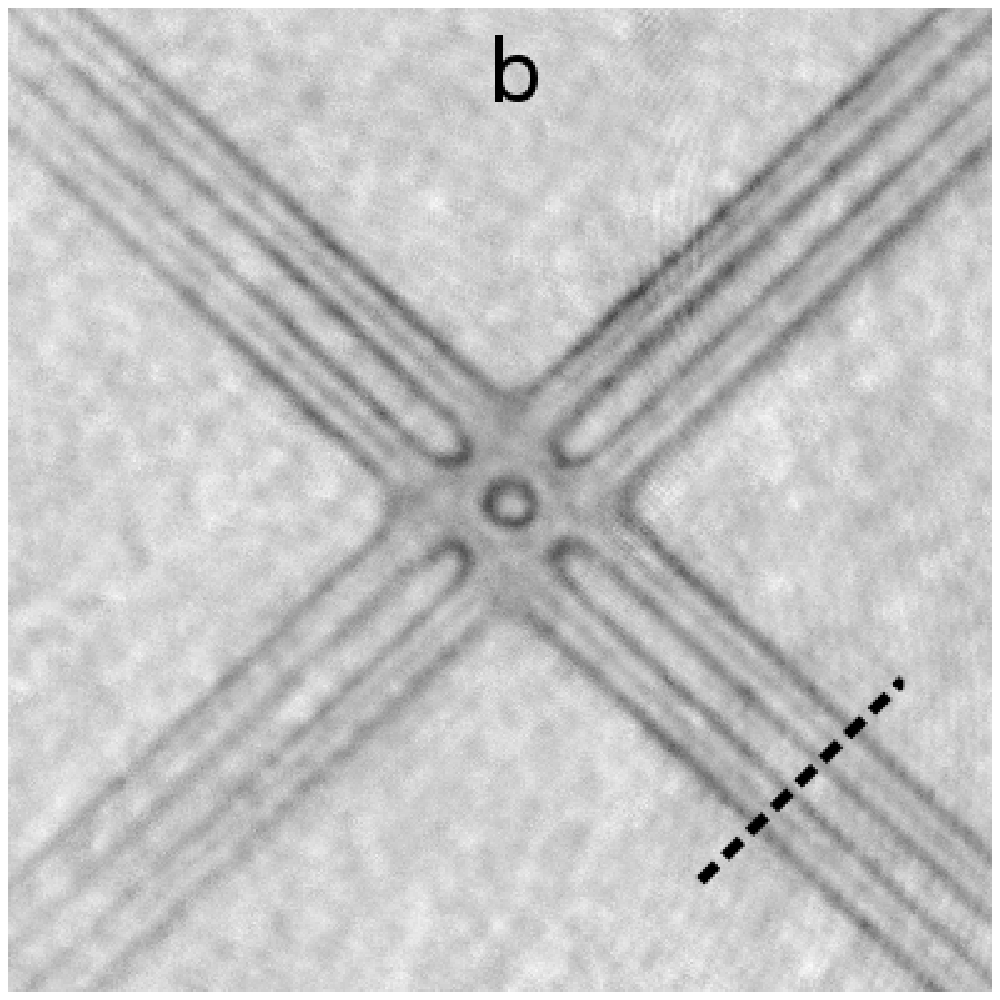}
\includegraphics[width=2.78cm]{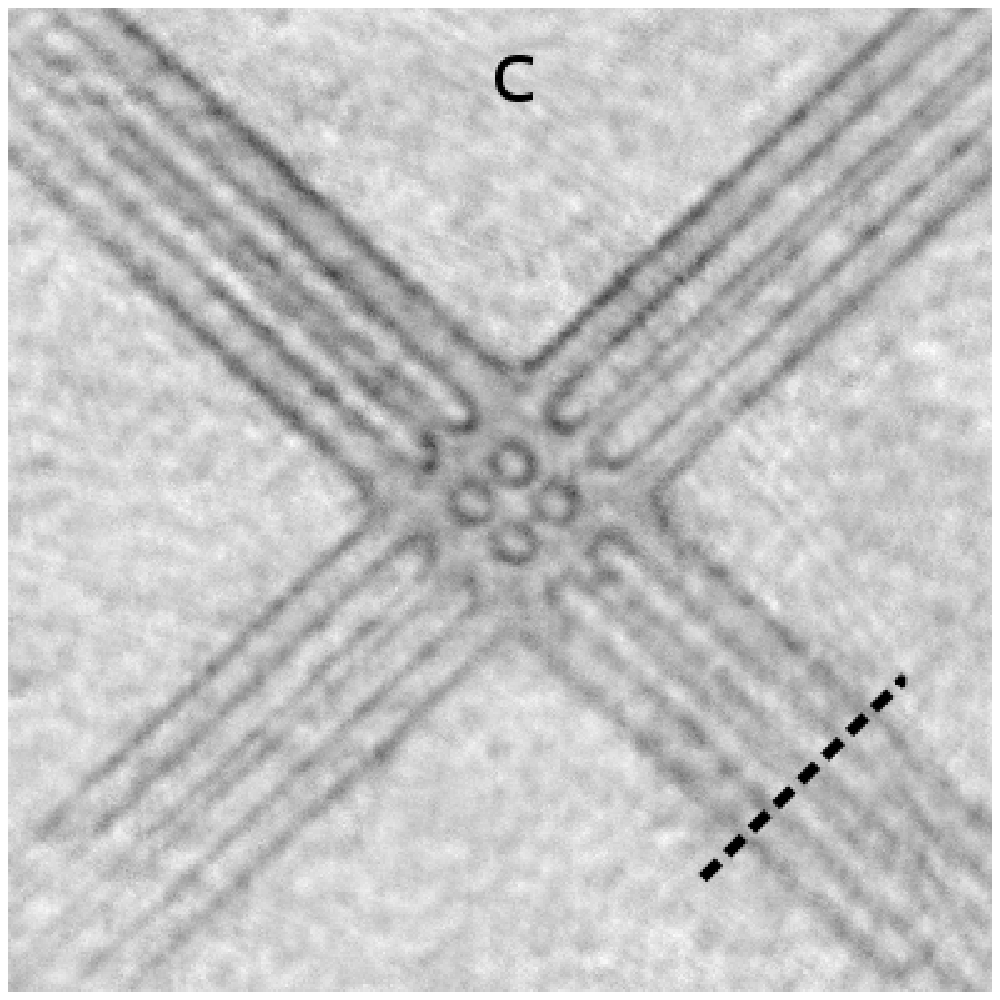}

\includegraphics[width=2.78cm]{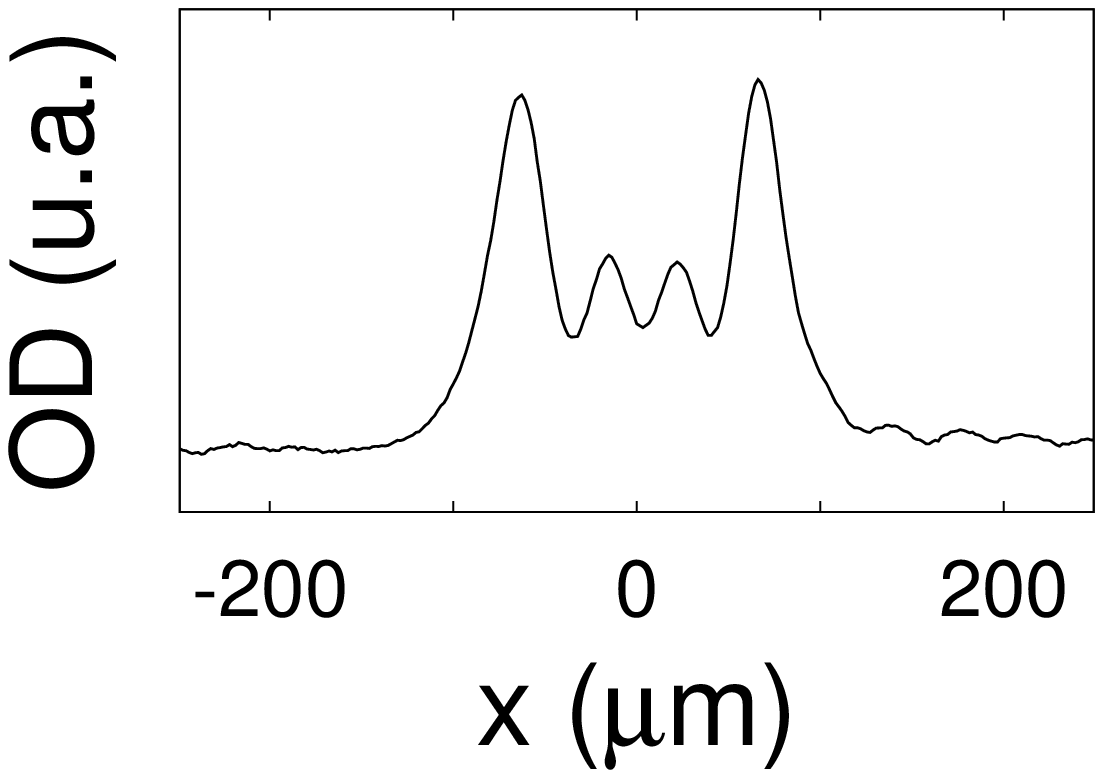}
\includegraphics[width=2.78cm]{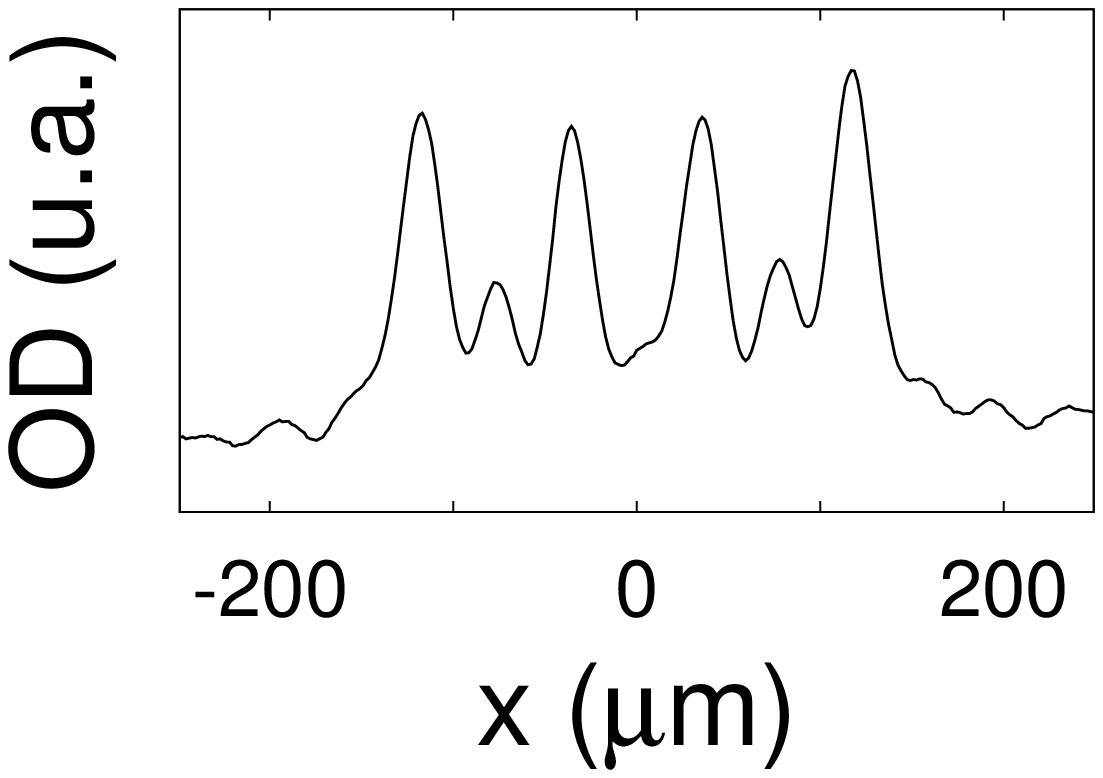}
\includegraphics[width=2.78cm]{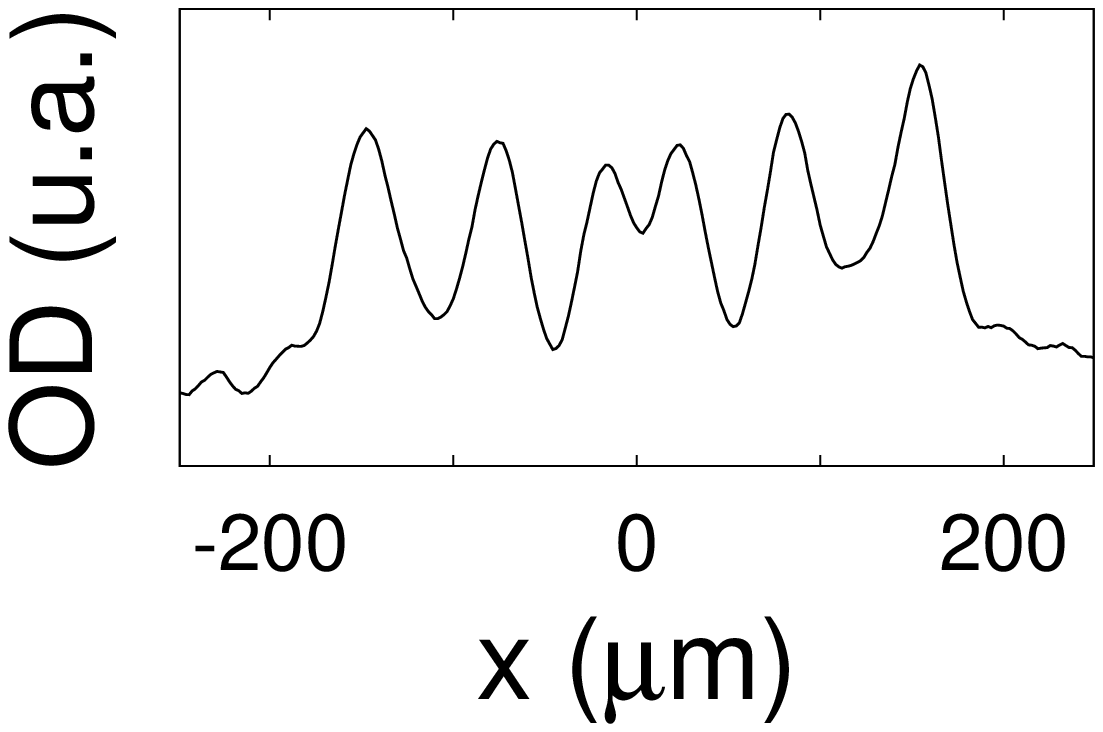}

\caption{\label{fig:tomography} (Top) Absorption images with the 1560 nm laser locked to the TEM$_{00}$
(a), TEM$_{10}$ (b), and TEM$_{20}$ (c) mode of the cavity, with the detuning of the probe light set to $5
\,\Gamma$ to the red of the transition. (Bottom) Integral optical density obtained by projecting the upper
images on the 45$^{\circ}$ dashed line crossing one arm of the cavity.}
\end{figure}

\begin{figure}[t]
\centerline{\includegraphics[width=9.0cm]{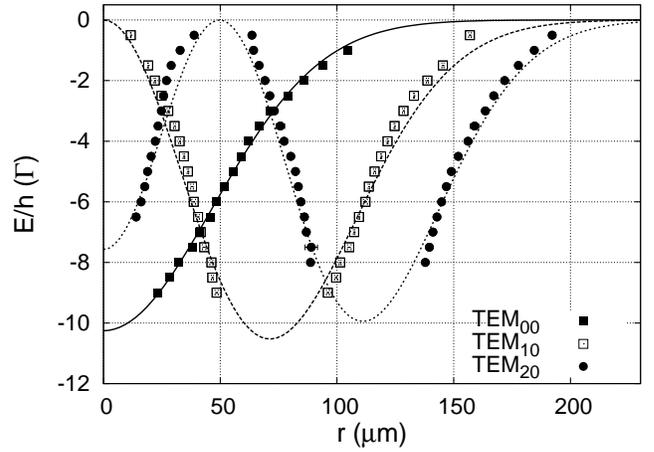}}
\caption{\label{fig:tem_profile} Optical potential depth induced by the laser at 1560 nm locked to the
first three transversal modes of the cavity. Each series of points is fitted with the corresponding
Hermite-Gaussian mode.}
\end{figure}

To determine the geometry and depth of the optical potential, the absorption of a probe beam was imaged at
different values of its detuning $\delta$ with respect to the $D_2$ line. The probe interacts with atoms
having potential energy $U(\mathbf{r})=\hbar \delta / (47.5-1)$ in the ground level; the width in energy of
the atomic class addressed is set by the linewidth of the probe transition ($\delta E = \hbar \Gamma /
(47.5-1)=k_B \times 6.3~\mu$K), since the linewidth of the probe laser is much smaller than $\Gamma$. If
the local potential energy gradient is known, $\delta E$ can be converted into spatial spread. A high ratio
between the maximum ground state energy shift in each cavity arm and $\delta E$ allows to finely section
the optical potential, but reduces the number of atoms probed at each measurement. A good compromise was
reached with a ratio of about 10, and the power of the input beam was set accordingly. To increase the
atomic density in the region of interest, the MOT cloud was compressed before its release. After 5 ms the
optical potential was imaged at different depths by measuring the atomic absorption with a 50 $\mu$s
optical pulse containing both probe and repumper radiation. Typical images are reported in Fig.
\ref{fig:tomography}, where the probe is detuned by $5 \, \Gamma$ with respect to the atomic transition and
the 1560 nm laser is locked to the first three transversal cavity modes. By fitting the projected optical
density signal reported in the lower row of Fig. \ref{fig:tomography} the position of the isopotential
lines was obtained. The smaller peaks in the projections of the TEM$_{00}$ and TEM$_{10}$ modes are due to
a second less intense probe component, frequency red shifted with respect to the main one: it shows that
the optical potential can be probed simultaneously at different depths. The measurement was repeated for
different values of the probe detuning as shown in Fig. \ref{fig:tem_profile}. The waist of the cavity was
obtained by fitting each series of data with the appropriate Hermite-Gauss mode: the result was 97(1),
101(1), and 99.8(4) $\mu$m for the TEM$_{00}$, TEM$_{10}$, and TEM$_{20}$ respectively. In the case of the
TEM$_{00}$ mode, the potential depth of the 5S$_{1/2}$ levels for each cavity arm is $k_B \times 60~\mu$K,
which corresponds to a power of 8.0 W in the cavity. Considering the power of the input beam and the
coupling efficiency, it means a cavity gain factor of 160. Using the full available power, the peak power
in the resonator reaches about 200 W, which means an optical potential depth per arm of $k_B \times 1.4$
mK. The resulting curvatures at the center of the potential dimple are 1.2 kHz along the two cavity arms,
and 1.6 kHz in the vertical direction. The good degree of vertical alignment for the two beams was
confirmed by the depth of the crossing region, which was measured to be twice the one--arm depth within
10\%.

In conclusion, we have frequency locked a fiber telecom laser to different transverse modes of a folded
optical cavity. The absorption of cold rubidium atoms, strongly modified by the light shift induced by
the 1560 nm radiation, was used to align the optical resonator and map \textit{in situ} the optical
potential. The radiation field obtained will be exploited to optically trap and evaporatively cool neutral
atoms to achieve a BEC directly in the cavity.

We thank J.P. Aoustin, A. G{\'e}rard, F. Moron, L. Pelay, and A. Villing for technical assistance, M.
Prevedelli and G. Santarelli for valuable advice on the laser lock. We acknowledge funding support
from DGA, IFRAF, EUROQUASAR, and FINAQS. The work of A. B. was supported by an European IEF Grant.

\bibliographystyle{osajnl}

\end{document}